\documentclass{article}
\usepackage{graphicx}
\usepackage{amsmath}
\usepackage{amsfonts}
\usepackage{amssymb}

\begin{document}
\ \ \ \ \ \bigskip

\bigskip

\begin{center}

\textbf{ALTERNATIVE STRUCTURES AND BIHAMILTONIAN SYSTEMS}

\vskip0.5cm G.Marmo$^{1,3}$,G.Morandi$^{2}$,A.Simoni$^{1,3}$ and
F.Ventriglia$^{1,4} $\\[0pt]{\small \emph{$^{1}$ Dipartimento di Scienze
Fisiche, Universita' di Napoli Federico II. \newline Complesso
Universitario di
Monte S.Angelo, Napoli, Italy}}\\[0pt]{\small \emph{$^{2}$ Dipartimento di
Fisica,Universita' di Bologna. INFM and INFN.\newline 6/2 v.le
B.Pichat. I-40127
Bologna, Italy}} \\[0pt]{\small \emph{$^{3}$ INFN, Sezione di di Napoli,}
} \\[0pt]{\small \emph{$^{4}$ INFM, UdR di Napoli.}}
{\small \emph{E-Mail:ventriglia@na.infn.it }}

%
\abstract
{In the study of bi-Hamiltonian systems (both classical and quantum) one starts with a given
dynamics and looks for all alternative Hamiltonian descriptions it admits.\\In this paper we start with two
compatible Hermitian structures (the quantum analog of two compatible classical Poisson brackets) and look for
all the dynamical systems which turn out to be bi-Hamiltonian with respect to them.}
\newpage
\end{center}

\bigskip

\bigskip

\section{\textbf{Introduction.}}

It is by now well known that the general structures of classical and quantum
systems are not essentially different. When considered as abstract dynamical
systems on infinite or finite dimensional vector spaces of states, both are
''Hamiltonian vector fields'' when considered in the Schr\"{o}dinger picture,
and both are ''inner derivations'' on the Lie algebra of observables with
respect to the Poisson brackets and the commutator brackets
respectively\cite{S66}. Moreover, in some appropriate limit, quantum mechanics
should reproduce classical mechanics. From this point of view, it is a natural
question to ask which alternative quantum descriptions of a given quantum
system would reproduce the alternative classical descriptions known as
bi-Hamiltonian descriptions of integrable systems \cite{biham}. This question
has been addressed recently in several collaborations involving some of us in
different combinations \cite{inv.us}.

When we consider composite systems and interactions among them it is
interesting to analyze to what extent these alternative quantum descriptions
survive. This paper is a preliminary step in this direction.

The specific problem we would like to address to can then be stated as follows:

\textit{''How many different quantum systems may, at the same time, admit of
bihamiltonian descriptions with respect to the same alternative
structures?''}

We will show that for generic, compatible, alternative structures
the different admissible quantum systems are pairwise commuting;
moreover, in finite dimensions, they close on a maximal torus.

To tackle the stated problem \cite{DMS}, we construct a quantum
system out of a given Hermitian structure and then we consider
compatible systems out of two Hermitian structures. The ''quantum
systems'' associated with a given Hermitian structure corresponds
to the infinitesimal generator of the ''phase group''. The
compatibility condition amounts to the requirement of
compatibility (commutativity) of the ''phase groups'' with respect
to both Hermitian structures.

All of our considerations will be carried over on finite-dimensional vector
spaces, and only in the final section we shall consider the extension of our
results to infinite dimensional Hilbert spaces.

\bigskip

\section{\textbf{Hermitian structures on }$R^{2n}$.}

In view of our interest in quantum systems, we will focus here our attention
on linear systems. Also, the tensorial structures that will be descried below
will be assumed to be represented by constant matrices. We will consider, to
start with, three relevant tensor structures that can be introduced in
$R^{2n}$, namely a \textbf{metric tensor} of the form:
\begin{equation}
g=g_{jk}dx^{j}\otimes dx^{k}%
\end{equation}
a \textbf{symplectic structure}:
\begin{equation}
\omega=\omega_{jk}dx^{j}\wedge dx^{k}%
\end{equation}
and a \textbf{complex structure} $J$, i.e. a $(1,1)$-type tensor satisfying:
\begin{equation}
J^{2}=-\mathbb{I}%
\end{equation}
We will be interested in the case in which the above structures are
\textbf{admissible}. By this we mean the following:

i) Suppose we are given $g$ and $J$. We will say that they are admissible, or
that $g$ is ''Hermitian'' iff:
\begin{equation}
g(Jx,Jy)=g(x,y)\text{ \ }\forall x,y
\end{equation}
We will always assume this to be the case.

Now, by virtue of $J^{2}=-\mathbb{I}$: $g(x,Jy)=-g(J(Jx),Jy)=-g(Jx,y).$
Hence:
\begin{equation}
g(Jx,y)+g(x,Jy)=0
\end{equation}

Notice that the previous two equations imply that $J$ \ generates
\textit{finite} as well as \textit{infinitesimal} rotations at the same time.
Moreover, Eq.$(5)$ implies that the $(0,2)$-type tensor:
\begin{equation}
\omega=:g(J.,.)
\end{equation}
is a symplectic form on $R^{2n}$ and that:
\begin{equation}
J=g^{-1}\circ\omega
\end{equation}
By proceeding as before one can prove that:
\begin{equation}
\omega(Jx,Jy)=\omega(x,y)
\end{equation}
and:
\begin{equation}
\omega(Jx,y)+\omega(x,Jy)=0
\end{equation}

$J$ will generate then both finite and infinitesimal symplectic transformations.

ii) Alternatively, one could start from $g$ and $\omega$ and say that they are
admissible iff: $J=g^{-1}\circ\omega$ satisfies: $J^{2}=-\mathbb{I}$. Written
explicitly in components this condition reads:
\begin{equation}
g^{jk}\omega_{kl}g^{lm}\omega_{mn}=-\delta_{n}^{j}%
\end{equation}

\textbf{Remarks.}

i) \textbf{ }Notice that if the condition $(4)$ does not hold, we can always
build a Hermitian structure out of a given $g$ by substituting it with the
symmetrized metric tensor:%
\begin{equation}
g_{s}(.,.)=:\frac{1}{2}\{g(J.,J.)+g(.,.)\}.
\end{equation}
which will be positive and nondegenerate if $g$ is.

Quite similarly\cite{Abra}, if the condition $(10)$ does not hold, then
Riesz's theorem tells us that there exists a nonsingular linear operator $A$
such that:%
\begin{equation}
\omega(x,y)=g(Ax,y)
\end{equation}
and the antisymmetry of $\omega$ implies:%
\begin{equation}
g(Ax,y)=-g(x,Ay)
\end{equation}
i.e. that $A$ is skew-hermitian: $A^{\dagger}=-A$, and: $-A^{2}>0$. Let then
$P$ be a (symmetric) nonnegative square root of $A$. $P$ will be injective,
and so $P^{-1}$ will be well defined\footnote{n the infinite-dimensional case,
it can be proved \cite{Abra} that $A$ is bounded and injective, and that $P$
is also injective and densely defined, so that $P^{-1}$ is well defined in the
infinite-dimensional case as well.}. We define then: $J=:AP^{-1}$ and:%
\begin{equation}
g_{\omega}(.,.)=:g(P(.),.)
\end{equation}

Therefore:%
\begin{equation}
\omega(x,y)=g(Ax,y)=g_{\omega}(Jx,y)
\end{equation}
and: $J^{\dagger}=-J,$ $J^{2}=-\mathbb{I}$ . The triple $(g_{\omega}%
,J,\omega)$ will be then an admissible triple, Eq.$(5)$ will hold true for
$g_{\omega}$ and, moreover:%
\begin{equation}
g_{\omega}(Jx,Jy)=g(Ax,Jy)=-g(AJx,y)=g_{\omega}(x,y)
\end{equation}
and Eq. $(4)$ will be satisfied as well.

ii) \ The adjoint $A^{\dagger}$ of any linear operator $A$ w.r.t. a metric
tensor $g$ is defined by the standard relation:%
\begin{equation}
g(A^{\dagger}x,y)=:g(x,Ay)
\end{equation}
and we can read Eq. $(5)$ as saying that the complex structure $J$ is
skew-adjoint w.r.t. the metric tensor $g.$

Although it may seem elementary, it is worth stressing here that, despite the
fact that we are working in a real vector space, the adjoint of $A$ does
\textbf{not} coincide with the transpose $A^{T}$ for a general $g$. Indeed,
spelling out explicitly Eq. $(17)$ in terms of matrices leads to:%
\begin{equation}
A^{\dagger}=g^{-1}A^{T}g
\end{equation}
and therefore, even for real matrices: $A^{\dagger}=A^{T}$ only if the metric
is standard Euclidean one and, in general, symmetric matrices need not be self-adjoint.

\bigskip

A linear structure on $R^{2n}$ is associated with a given dilation
(or Liouville) vector field $\Delta$. Given then a linear
structure on $\mathbb{R}^{2n}$, we can associate with every matrix
$\mathbb{\ A\equiv}||A^{i}$ $_{j}||\mathbb{\in}\
\frak{gl}(2n,\mathbb{R})$ both a
$(1,1)$-type tensor field:%
\begin{equation}
T_{\mathbb{A}}=A^{i}_{j}dx^{j}\otimes\frac{\partial}{\partial x^{i}}%
\end{equation}
and a linear vector field:%
\begin{equation}
X_{\mathbb{A}}=A^{i}_{j}x^{j}\frac{\partial}{\partial x^{i}}%
\end{equation}
The two are connected by:%
\begin{equation}
T_{\mathbb{A}}(\Delta)=X_{\mathbb{A}}%
\end{equation}
and are both homogeneous of degree zero, i.e.:%
\begin{equation}
L_{\Delta}X_{\mathbb{A}}=L_{\Delta}T_{\mathbb{A}}=0
\end{equation}

The correspondence: $\mathbb{A}\rightarrow T_{\mathbb{A}}$ is \ (full)
associative algebra and Lie algebra isomorphism. The correspondence:
$\mathbb{A}\rightarrow X_{\mathbb{A}}$ is instead only a Lie algebra
(anti)isomorphism, i.e.:%
\begin{equation}
T_{\mathbb{A}}\circ T_{\mathbb{B}}=T_{\mathbb{AB}}%
\end{equation}
while:%
\begin{equation}
\lbrack X_{\mathbb{A}},X_{\mathbb{B}}]=-X_{[\mathbb{A},\mathbb{B}]}%
\end{equation}
Moreover, for any $\mathbb{A},\mathbb{B}\in$ $\frak{gl}(2n,\mathbb{R})$:%
\begin{equation}
L_{X_{\mathbb{A}}}T_{\mathbb{B}}=-T_{_{[\mathbb{A},\mathbb{B}]}}%
\end{equation}

This implies that all statements that can be proved at the level of vector
fields and/or at that of $(1,1)$ tensors can be rephrased into equivalent
statements in terms of the corresponding representative matrices, and
viceversa. That is why we will work mostly directly with the representative
matrices in what follows.

\bigskip

Out of the Liouville field and the metric tensor we can construct the
quadratic function:
\begin{equation}
\mathsf{g}=\frac{1}{2}g(\Delta,\Delta)
\end{equation}
and the associated Hamiltonian vector field $\Gamma$ via:
\begin{equation}
i_{\Gamma}\omega=-d\mathsf{g}%
\end{equation}
In a given coordinate system $(x^{1},...,x^{2n})$: $\Delta=x^{i}%
\partial/\partial x^{i},\Gamma=\Gamma^{k}\partial/\partial x^{k}$ and,
explicitly: $\omega_{jk}\Gamma^{k}=-g_{jk}x^{k},$ implying:
$g^{lj}\omega _{jk}\Gamma^{k}=-x^{l}$ i.e.:
$\Gamma^{k}=J^{k}_{l}x^{l}$, or:
\begin{equation}
J(\Gamma)=-\Delta\Leftrightarrow\Gamma=J(\Delta)
\end{equation}
$\Gamma$ will preserve both $g$ and $\omega$, and hence $J$ (more generally,
it will preserve any third structure if it preserves the other two):
\begin{equation}
L_{\Gamma}\omega=L_{\Gamma}g=L_{\Gamma}J=0
\end{equation}

Given a metric tensor and an admissible symplectic form, an
\textbf{hermitian structure} on $\mathbb{R}^{2n}$ is a map: $h:$
$\mathbb{R}^{2n}\times
\mathbb{R}^{2n}\rightarrow\mathbb{R}^{2}$defined as:
$$h:(x,y)\mapsto (g(x,y),\omega(x,y)).\nonumber$$ Equivalently (and having in
mind quantum systems) one can exploit the fact that $R^{2n}$ can
be given a complex vector space structure by defining, for
$z=\alpha+i\beta\in\mathbb{C}$ \ and $x\in
\mathbb{R}^{2n}$:%
\begin{equation}
z\cdot x=(\alpha+i\beta)x=:\alpha x+J\beta x \, ;
\end{equation}
then $h$ will become  an hermitian scalar product, linear in the
second factor, on \ a complex vector space, and we can write now:
\begin{equation}
h(x,y)=g(x,y)+ig(Jx,y)
\end{equation}
and of course a statement equivalent to the previous ones will be:
\begin{equation}
L_{\Gamma}h=0
\end{equation}
The vector field $\Gamma$ will be therefore a generator of the unitary group
on $C^{n}$, and will be an instance of what we will call a \textbf{quantum
system}. More generally, a quantum system will be any linear vector field:
\begin{equation}
\Gamma_{\mathbb{A}}=A^{j}_{k}x^{k}\frac{\partial}{\partial x^{j}}%
\end{equation}
associated with a matrix: $\mathbb{A}=||A^{j}$$_{k}||$ that preserves both $g$
and $\omega$ or, equivalently, $h$:
\begin{equation}
L_{\Gamma_{\mathbb{A}}}h=0
\end{equation}

This defining requirement on $\Gamma_{\mathbb{A}}$ implies that the matrix
$\mathbb{A}$ in the description of $\Gamma_{\mathbb{A}}$ be at the same time
an infinitesimal generator of a realization of the symplectic group $Sp(n)$
and of a realization of the rotation group $SO(2n)$. The intersection of these
two Lie algebras yields a realization of the Lie algebra of the unitary group.

\bigskip

\bigskip

\section{\textbf{\ Bihamiltonian descriptions.}}

\bigskip

Consider now two different Hermitian structures on $R^{2n}$: $h_{1}%
=g_{1}+i\omega_{1}$ and: $h_{2}=g_{2}+i\omega_{2}$, with the associated
quadratic functions: \textsf{g}$_{1}=g_{1}(\Delta,\Delta)$, \textsf{g}%
$_{2}=g_{2}(\Delta,\Delta)$ and Hamiltonian vector fields $\Gamma_{1}$and
$\Gamma_{2}$. Then: \bigskip

\textbf{Definition: }$h_{1}$and $h_{2}$ will be said to be \textbf{compatible}
iff:
\begin{equation}
L_{\Gamma_{1}}h_{2}=L_{\Gamma_{2}}h_{1}=0
\end{equation}

This will imply of course:
\begin{equation}
L_{\Gamma_{1}}\omega_{2}=L_{\Gamma_{1}}g_{2}=0
\end{equation}
separately, and similar equations with the indices interchanged.

\bigskip\textbf{Remark. }Notice that, if: $\omega=\frac{1}{2}\omega_{ij}%
dx^{i}\wedge dx^{j}$ is a constant symplectic structure and:
$X=A^{i}_{j}x^{j}\partial/\partial x^{i}$ is a linear vector
field, then the condition: $L_{X}\omega=0$ can be written in terms
of the representative
matrices as the requirement that the matrix $\omega A$ be symmetric, i.e.:%
\begin{equation}
\omega A=(\omega A)^{T}\Leftrightarrow\omega A+A^{T}\omega=0
\end{equation}
while the condition: $L_{X}g=0$ implies that the matrix $gA$ be
skew-symmetric, i.e.:%
\begin{equation}
gA+(gA)^{T}=gA+A^{T}g=0
\end{equation}

\bigskip

Notice now that, from: $i_{\Gamma_{2}}\omega_{2}=-d\mathsf{g}_{2}$
and: $L_{\Gamma_{1}}g_{2}=0$ we obtain:
$$0=L_{\Gamma_{1}}(i_{\Gamma_{2}}\omega
_{2})=:L_{\Gamma_{1}}\omega_{2}(\Gamma_{2},.)=(L_{\Gamma_{1}}\omega
_{2})(\Gamma_{2},.)+\omega_{2}([\Gamma_{1},\Gamma_{2}],.)\nonumber$$
and as $L_{\Gamma_{1}}\omega_{2}=0$:
\begin{equation}
i_{[\Gamma_{1},\Gamma_{2}]}\omega_{2}=0
\end{equation}
and similarly for $\omega_{1}.$ As neither $\omega_{1}$ nor $\omega_{2}$ is
degenerate, this implies that $\Gamma_{1}$ and $\Gamma_{2}$ commute, i.e.
that:
\begin{equation}
\lbrack\Gamma_{1},\Gamma_{2}]=0
\end{equation}
Moreover, remembering that, given a symplectic structure $\omega$, the Poisson
bracket of any two functions $f$ and $g$, $\{f,g\}$ is defined as:
$\{f,g\}=:\omega(X_{g},X_{f})$, where $X_{f}$ and $X_{g}$ are the Hamiltonian
vector fields associated with $f$ and $g$ respectively, we find:
$0=L_{\Gamma_{1}}\mathsf{g}_{2}=d\mathsf{g}_{2}(\Gamma_{1})=$ $=-\omega
_{2}(\Gamma_{2},\Gamma_{1})$. Hence we find:%

\begin{equation}
\{\mathsf{g}_{1},\mathsf{g}_{2}\}_{2}=0
\end{equation}
where $\{.,.\}_{2}$ is the Poisson bracket associated with
$\omega_{2}$ and similarly for $\omega_{1}$.

\bigskip

\textbf{Remark.}The four real conditions: $\{\mathsf{g}_{1},\mathsf{g}%
_{2}\}_{1,2}=0$ and: $L_{\Gamma_{1}}\omega_{2}=L_{\Gamma_{2}}\omega_{1}=0$ are
actually equivalent to those stated in complex form in Eq.$(35)$.

\bigskip

Remembering what has already been said about the fact that statements
concerning linear vector fields translate into equivalent statements for the
$(1,1)$-type tensor fields having the same representative matrices, and
recalling that the defining matrices of $\ \Gamma_{1}$ and $\Gamma_{2}$ are
precisely those of the corresponding complex structures, we see at once that:
\begin{equation}
\lbrack\Gamma_{1},\Gamma_{2}]=0\Leftrightarrow\lbrack J_{1},J_{2}]=0
\end{equation}
i.e. that the two complex structures will commute as well.

In general, given any two $(0,2)$ (or $(2,0)$) tensor fields $h$ and $g$ one
(at least) of which, say $h$, is invertible, the composite tensor $h^{-1}\circ
g$ will be a $(1,1)$-type tensor. Then, out of the two compatible structures
we can build up the two $(1,1)$-type tensor fields:%
\begin{equation}
G=g_{1}^{-1}\circ g_{2}%
\end{equation}
and:%
\begin{equation}
T=\omega_{1}^{-1}\circ\omega_{2}%
\end{equation}

Actually one can prove at once that the two are related, and indeed direct
calculation proves that:
\begin{equation}
G=J_{1}\circ T\circ J_{2}^{-1}\equiv-J_{1}\circ T\circ J_{2}\Leftrightarrow
\text{ }T=-J_{1}\circ G\circ J_{2}%
\end{equation}
It turns out that $T$ (and hence $G$) commutes with both complex structures i.e.:%

\begin{equation}
\lbrack G,J_{a}]=[T,J_{a}]=0,\text{ }a=1,2
\end{equation}

This follows from the fact that both $G$ and $T$ are $\Gamma$-invariant, i.e.:%
\begin{equation}
L_{\Gamma_{1,2}}G=L_{\Gamma_{1,2}}T=0
\end{equation}
and from Eq. $(25)$.

It follows also from Eqns. $(25)$ and $(26)$ that $G$ and $T$ commute, i.e.:
\begin{equation}
\lbrack G,T]=0
\end{equation}

Moreover, $G$ enjoys the property that:
\begin{equation}
g_{a}(Gx,y)=g_{a}(x,Gy),\text{ }a=1,2
\end{equation}
Indeed one can prove by direct calculation that:
\begin{equation}
g_{1}(Gx,y)=g_{1}(x,Gy)=g_{2}(x,y)
\end{equation}
while: $\ $%
\begin{equation}
g_{2}(Gx,y)=g_{2}(x,Gy)=g_{1}^{-1}(g_{2}(x,.),g_{2}(y,.))
\end{equation}
and this completes the proof. Eq.$(49)$ can be seen as saying that $G$ is
''self-adjoint'' w.r.t. both metrics.

Notice that the derivation of this result does not require the compatibility
condition to hold. If the latter is assumed, however, one can prove also that
$T$ is self-adjoint w.r.t. both metrics, and that both $J_{1}$and $J_{2}$ are
instead skew-adjoint w.r.t. both structures, i.e. that:%
\begin{equation}
g_{a}(Tx,y)=g_{a}(x,Ty),\text{ \ }a=1,2
\end{equation}
and that:%
\begin{equation}
g_{1}(x,J_{2}y)+g_{1}(J_{2}x,y)=0\text{ \ }\forall x,y
\end{equation}
with a similar equation with the indices interchanged.

Indeed, from, e.g.: $L_{\Gamma_{1}}\omega_{2}=0$ we obtain, in terms of the
representative matrices and using Eq.(37) and: $J_{1}=g_{1}^{-1}\omega_{1}$:
\begin{equation}
\omega_{2}g_{1}^{-1}\omega_{1}=\omega_{1}g_{1}^{-1}\omega_{2}\Leftrightarrow
\omega_{2}\omega_{1}^{-1}g_{1}=g_{1}\omega_{1}^{-1}\omega_{2}%
\end{equation}
Remembering the definition of $T$, this is equivalent to: $g_{1}T=(g_{1}%
T)^{T}$, and this leads to:%
\begin{equation}
T=g_{1}^{-1}T^{T}g_{1}=(T^{\dagger})_{1}%
\end{equation}
where $(T^{\dagger})_{1}$ is the adjoint of $T$ w.r.t. $g_{1}$. Interchanging
indices, one can prove that: $(T^{\dagger})_{2}=T$ as well.

Concerning the $J$'s (that have already been proved to be skew-adjoint w.r.t.
the respective metric tensors), consider, e.g. $$(J_{1}^{\dagger})_{2}%
=:g_{2}^{-1}J_{1}^{T}g_{2}
=-g_{2}^{-1}g_{1}J_{1}^{T}g_{1}^{-1}g_{2}=-G^{-1}J_{1}G=-J_{1}\nonumber$$
as $G$ and the $J$'s commute. A similar result holds of course for
$J_{2}$.$\blacksquare$

Summarizing what has been proved up to now, we have found that $G,T,J_{1}$and
$J_{2}$ are a set of mutually commuting linear operators. $G$ and $T$ are
self-adjoint, while $J_{1}$ and $J_{2}$ are skew-adjoint, w.r.t. both metric tensors.

If we now diagonalize $G$, the $2n$-dimensional vector space $V=\mathbb{R}%
^{2n}$ will split into a direct sum of eigenspaces: $V=\oplus_{k}%
V_{\lambda_{k}}$, where the $\lambda_{k}$'s ($k=1,...r\leq2n$) are the
distinct eigenvalues of $G$. According to what has just been proved, the sum
will be an orthogonal sum w.r.t. both metrics, and, in $V_{\lambda_{k}}$,
$G=\lambda_{k}\mathbb{I}_{k}$, with $\mathbb{I}_{k}$ the identity matrix in
$V_{\lambda_{k}}$. Assuming compatibility, $T$ will commute with $G$ and will
be self-adjoint. Therefore we will get a further orthogonal decomposition of
each $V_{\lambda_{k}}$ of the form:%
\begin{equation}
V_{\lambda_{k}}=\bigoplus_{r}W_{\lambda_{k},\mu_{k,r}}%
\end{equation}
where the $\mu_{k,r}$'s are the (distinct) eigenvalues of $T$ in
$V_{\lambda_{k}}$.The complex structures commute in turn with both $G$ and
$T$. Therefore they will leave each one of the $W_{\lambda_{k},\mu_{k,r}}$'s invariant.

Now we can reconstruct, using the $g$'s and the $J$'s, the two
symplectic structures. They will be block-diagonal in the
decomposition $(47)$ of $V$,
and on each one of the $W_{\lambda_{k},\mu_{k,r}}$'s they will be of the form:%
\begin{equation}%
\begin{array}
[c]{c}%
g_{1}=\lambda_{k}g_{2}\\
\omega_{1}=\mu_{k,r}\omega_{2}%
\end{array}
\end{equation}

Therefore, in the same subspaces: $J_{1}=g_{1}^{-1}\omega_{1}=\frac{\mu_{k,r}%
}{\lambda_{k}}J_{2}$ . It follows from: $J_{1}^{2}=$
$J_{2}^{2}=-1$ that: $(\frac{\mu_{k,r}}{\lambda_{k}})^{2}=1,$
whence: $\ \mu_{k,r}=\pm\lambda_{k}$ (and $\lambda_{k}>0$ ). The
index $r$ can then assume only two values, corresponding to
$\pm\lambda_{k}$ and at most $V_{\lambda_{k}}$ will have the
decomposition of $V_{\lambda_{k}}$ into the orthogonal sum:
$V_{\lambda_{k}}=$ $W_{\lambda_{k},\lambda_{k}}\oplus
W_{\lambda_{k},-\lambda_{k}}$. All in all, what we have proved is
the following:
\bigskip

\textbf{Lemma.} If the two hermitian structures $h_{1}=(g_{1},\omega_{1})$ and
$h_{2}=(g_{2},\omega_{2})$\footnote{Coming, of course, from admissible triples
$(g_{1},\omega_{1},J_{1})$ and \ $(g_{2},\omega_{2},J_{2})$.} are compatible,
then the vector space $V\approx\mathbb{R}^{2n}$ will decompose into the
(double) orthogonal sum:%
\begin{equation}
\underset{k=1,...,r;\text{ \ }\alpha=\pm}{ \bigoplus }W_{\lambda
_{k},\alpha\lambda_{k}}%
\end{equation}
where the index $k=1,...,r\leq2n$ labels the eigenspaces of the
$(1,1)$-type tensor: $G=g_{1}^{-1}\circ g_{2}$ corresponding to
its distinct eigenvalues $\lambda_{k}>0$, while:
$T=\omega_{1}^{-1}\circ\omega_{2}$ will be diagonal (with
eigenvalues $\pm\lambda_{k})$ on the
$W_{\lambda_{k},\pm\lambda_{k}}$'s, on each one of which:

\begin{equation}%
\begin{array}
[c]{c}%
g_{1}=\lambda_{k}g_{2}\text{ \ }\\
\omega_{1}=\pm\lambda_{k}\omega_{2}\text{ \ }\\
J_{1}=\pm J_{2}%
\end{array}
\end{equation}

As neither symplectic form is degenerate, the dimension of each one of the
$W_{\lambda_{k},\pm\lambda_{k}}$'s will be necessarily even.$\blacksquare$

\bigskip

Now we can further qualify \ and strengthen the compatibility condition by
stating\ the following:

\bigskip

\textbf{Definition: }Two (compatible) Hermitian structures will be said to be
\textbf{in a generic position} iff the eigenvalues of $G$ and $T$ have minimum
(i.e. double) degeneracy.

\bigskip

In general, two appropriate geometrical objects like two $(0,2)$ or
$(2,0)$-type \ tensor fields are said to be in a generic position if they can
be ''composed'' to yield a 1-1 tensor whose eigenvalues have minimum
degeneracy. For instance $g_{1}$ and $g_{2}$ are in a generic position if the
eigenvalues of $G=g_{1}^{-1}\circ g_{2}$ have minimum degeneracy, which
possibly depends on further conditions: when the compatibility is required,
this degeneracy is double. The results that we have just proved will imply
that each one of the $W_{\lambda_{k},\lambda_{k}},W_{\lambda_{k},-\lambda_{k}%
}$ will have the minimum possible dimension, that is two.

Denoting then by $E_{k}$ ($k=1,...,n$, now) these two-dimensional subspaces,
all that has been said up to now can be summarized in the following:

\bigskip

\textbf{Proposition: }If $h_{1}$ and $h_{2}$ are compatible and in a generic
position, then $\mathbb{R}^{2n}$ splits into a sum of $n$ mutually
''bi-orthogonal'' (i.e. orthogonal with respect to both metrics $g_{1}$ and
$g_{2}$ ) two-dimensional vector subspaces : $\mathbb{R}^{2n}=E_{1}\oplus
E_{2}\oplus....\oplus E_{n}$ . All the structures $g_{a},J_{a},\omega_{a}$
decompose accordingly into a direct sum of structures on these two-dimensional
subspaces, and on each one of the $E_{k}$'s they can be written as:%

\begin{equation}%
\begin{array}
[c]{cc}%
g_{1}|_{E_{k}}=\lambda_{k}(e_{1}^{\ast}\otimes e_{1}^{\ast}+e_{2}^{\ast
}\otimes e_{2}^{\ast});\ \lambda_{k}>0 & g_{2}|_{E_{k}}=\varrho_{k}%
\ g_{1}|_{E_{k}};\ \varrho_{k}>0\\
J_{1}|_{E_{k}}=(e_{2}e_{1}^{\ast}-e_{1}e_{2}^{\ast})\ \ \ \ \ \ \
\ \ &
J_{2}|_{E_{k}}=\pm J_{1}|_{E_{k}}%
\ \ \ \ \ \ \ \ \ \ \ \ \\
\omega_{1}|_{E_{k}}=\lambda_{k}(e_{1}^{\ast}\wedge e_{2}^{\ast}%
)\ \ \ \ \ \ \ \  & \omega_{2}|_{E_{k}}=\pm\varrho_{k}\,\omega_{1}|_{E_{k}%
}\ \ \ \ \ \ \ \ \ \ \ \
\end{array}
\end{equation}
where $e_{2}=J_{1}e_{1}$ , $e_{1}$ is any given vector in $E_{k}$
and the $e^{\ast}$'s are the dual basis of the $e^{\prime}$ 's
$\footnote{In other words on each subspace $g_{1}$ and $g_{2}$ are
proportional, while $J_{1}=$ $\pm J_{2}$ and accordingly
$\omega_{2}=\pm\varrho\omega_{1}.$}.\blacksquare$

Every linear vector field preserving both $h_{1}=(g_{1},\omega_{1})$ and
$h_{2}=(g_{2},\omega_{2})$ will have a representative matrix commuting with
those of $T$ \ and $G$ , and it will be block-diagonal in the same eigenspaces
$E_{k}$. Therefore, in the generic case, the analysis can be restricted to
each 2-dimensional subspace $\ E_{k}$ in which the vector field will preserve
both a symplectic structure and a positive-definite metric. Therefore it will
be in $sp(2)\cap SO(2)=U(1)$ and, on each $\ E_{k}$, it will represent a
harmonic oscillator with frequencies depending in general on the $V_{k}$'s .

Having discussed the general case, and to gather more insight into the problem
we are discussing here, we will describe now in full details the
two-dimensional case.

\bigskip

\textbf{A two-dimensional example.}

\bigskip

Starting from the observation that two quadratic forms\footnote{One of which
is assumed to be positive.} can always be diagonalized simultaneously (at the
price of using a non-orthogonal transformation, if necessary) we can assume
from start $g_{1}$ and $g_{2}$ to be of the form:%
\begin{equation}
g_{1=}\left|
\begin{array}
[c]{cc}%
\varrho_{1} & 0\\
0 & \varrho_{2}%
\end{array}
\right|
\end{equation}
and:%
\begin{equation}
g_{2=}\left|
\begin{array}
[c]{cc}%
\sigma_{1} & 0\\
0 & \sigma_{2}%
\end{array}
\right|
\end{equation}

\bigskip The more general $J$ such that $J^{2}=-1$ will be of the form:%
\begin{equation}
J=\left|
\begin{array}
[c]{cc}%
a & b\\
-\frac{(1+a^{2})}{b} & -a
\end{array}
\right|
\end{equation}
Compatibility with $g_{1}$ requires that $J$ be anti-hermitian (w.r.t. $g_{1}%
$), and this leads to:%
\begin{equation}
J=J_{1\pm}=\left|
\begin{array}
[c]{cc}%
0 & \pm\sqrt{\frac{\varrho_{2}}{\varrho_{1}}}\\
\mp\sqrt{\frac{\varrho_{1}}{\varrho_{2}}} & 0
\end{array}
\right|
\end{equation}
and similarly:%
\begin{equation}
J=J_{2\pm}=\left|
\begin{array}
[c]{cc}%
0 & \pm\sqrt{\frac{\sigma_{2}}{\sigma_{1}}}\\
\mp\sqrt{\frac{\sigma_{1}}{\sigma_{2}}} & 0
\end{array}
\right|
\end{equation}
from the requirement of admissibility with $g_{2}$.

\bigskip

As a consequence:
\begin{equation}
\omega=\omega_{1\pm}=\left|
\begin{array}
[c]{cc}%
0 & \pm\sqrt{\varrho_{2}\varrho_{1}}\\
\mp\sqrt{\varrho_{2}\varrho_{1}} & 0
\end{array}
\right|
\end{equation}
and:%
\begin{equation}
\omega=\omega_{2\pm}=\left|
\begin{array}
[c]{cc}%
0 & \pm\sqrt{\sigma_{2}\sigma_{1}}\\
\mp\sqrt{\sigma_{2}\sigma_{1}} & 0
\end{array}
\right|
\end{equation}
Now we have all the admissible structures, i.e. $\
(g_{1},\omega_{1\pm },J_{1\pm})$ and
$(g_{2},\omega_{2\pm},J_{2\pm})$.

Let's compute the invariance group for the first triple having
made a definite choice for the possible signs \ (say: $J=$
$J_{+}$). The group is easily seen
to be:%
\begin{equation}
O_{1}(t)=\cos(t)\mathbb{I}+\sin(t)J_{1}=\left|
\begin{array}
[c]{cc}%
\cos(t) & \sqrt{\frac{\varrho_{2}}{\varrho_{1}}}\sin(t)\\
-\sqrt{\frac{\varrho_{1}}{\varrho_{2}}}\sin(t) & \cos(t)
\end{array}
\right|
\end{equation}
while for the second triple we obtain:%
\begin{equation}
O_{2}(t)=\cos(t)\mathbb{I}+\sin(t)J_{2}=\left|
\begin{array}
[c]{cc}%
\cos(t) & \sqrt{\frac{\sigma_{2}}{\sigma_{1}}}\sin(t)\\
-\sqrt{\frac{\sigma_{1}}{\sigma_{2}}}\sin(t) & \cos(t)
\end{array}
\right|
\end{equation}
and in general we obtain two different realizations of $SO(2)$.

The two realizations have only a trivial intersection (coinciding
with the identity) if
$\rho_{2}/\rho_{1}\neq\sigma_{2}/\sigma_{1}$, and coincide when
$\rho_{2}/\rho_{1}=\sigma_{2}/\sigma_{1}$. The latter condition is
easily seen (by imposing e.g.: $[J_{1,2},T]=0$) to be precisely
the condition of compatibility of the two triples.

To conclude the discussion of the example, let's see what happens in the
complexified version of the previous discussion.

To begin with we have to define multiplication by complex numbers on
$\mathbb{R}^{2}$, thus making it a complex vector space, and this can be done
in two ways, namely as:%
\begin{equation}
(x+iy)\left|
\begin{array}
[c]{c}%
a\\
b
\end{array}
\right|  =(x\mathbb{I+}J_{1}y)\left|
\begin{array}
[c]{c}%
a\\
b
\end{array}
\right|
\end{equation}
or as:%
\begin{equation}
(x+iy)\left|
\begin{array}
[c]{c}%
a\\
b
\end{array}
\right|  =(x\mathbb{I+}J_{2}y)\left|
\begin{array}
[c]{c}%
a\\
b
\end{array}
\right|
\end{equation}

Correspondingly, we can introduce two different hermitian structures on
$\mathbb{R}^{2}$ \ as:%
\begin{equation}
(.,.)_{1}=g_{1}+i\omega_{1}%
\end{equation}
or as:%
\begin{equation}
(.,.)_{2}=g_{2}+i\omega_{2}%
\end{equation}
They are antilinear in the first factor and in each case the
corresponding multiplication by complex numbers must be used. The
$O_{1}(t)$ and $O_{2}(t)$ actions both coincide with the
multiplication of points of $\mathbb{R}^{2}$ by the complex
numbers $e^{it}$ (i.e. with different realizations of $U(1)$), but
the definition of multiplication by complex numbers is different
in the two cases.

\bigskip

Going back to the general case, we can make contact with \ the theory of
complete integrability of bi-hamiltonian system by observing that $T$ plays
here the role of \ a recursion operator\cite{rec2}. Indeed, we show now that
it generates a basis of vector fields preserving both the hermitian structures
$h_{a}$ given by:
\begin{equation}
\Gamma_{1},T\Gamma_{1},...,T^{n-1}\Gamma_{1}\ \ \ \ .
\end{equation}

To begin with, these fields preserve all the geometrical
structures, commute pairwise and are linearly independent. In fact
these properties follow from the observation that $T$, being a
constant 1-1 tensor, satisfies the Nijenhuis condition \cite{Ni} .
Therefore, for any vector field $X$:
\begin{equation}
L_{TX}T=TL_{X}T\
\end{equation}
that, $T$ \ being invertible, amounts to:%
\begin{equation}
L_{TX}=TL_{X}%
\end{equation}
So, $\forall k\in\mathbb{N}$:
\begin{equation}
L_{T^{k}\Gamma_{1}}=TL_{T^{k-1}\Gamma_{1}}=...=T^{k}L_{_{\Gamma_{1}}}%
\end{equation}
and
\begin{equation}
T^{k}L_{_{\Gamma_{1}}}\omega_{a}=0=T^{k}L_{_{\Gamma_{1}}}g_{a}\ \ \ \ ;
\end{equation}

Moreover, $\forall s\in\mathbb{N}$:
\begin{equation}
\lbrack T^{k+s}\Gamma_{1},T^{k}\Gamma_{1}]=L_{T^{k+s}\Gamma_{1}}T^{k}%
\Gamma_{1}=T^{s}L_{T^{k}\Gamma_{1}}T^{k}\Gamma_{1}=T^{s}[T^{k}\Gamma_{1}%
,T^{k}\Gamma_{1}]=0 .
\end{equation}

Besides, the assumption of minimal degeneracy of $T$ implies that the minimal
polynomial\cite{L} of $T$ \ be of degree $n$. Indeed, we have shown that the
diagonal form of $T$ is
\begin{equation}
T=\underset{k=1,...,n}{\bigoplus}\{\pm\rho_{k}\mathbb{I}_{k}\}
\end{equation}
where $\mathbb{I}_{k}$ is the identity on $V_{k}$ . Any linear combination
\begin{equation}
\sum\limits_{r=0}^{m}\alpha_{r}T^{r}=0\ \ \ ,\ \ \ m\leq n-1\ \ \ ,
\end{equation}
yields a linear system for the $\alpha_{r}$'s of $n$ equations in $m+1$
unknowns whose matrix of coefficients is of maximal rank and that, for
$m=n-1$, coincides with the full Vandermonde matrix of the $\rho_{k}$'s .

Then, we can conclude that the $n$ vector fields $T^{r}\Gamma_{1}%
,r=0,1,...,n-1$ form a basis.

\bigskip

\bigskip

\bigskip

\section{The infinite-dimensional case}

\bigskip

We now analyze the same kind of problems in the framework of Quantum
Mechanics, taking advantage from the experience and results we have obtained
in the previous Sections where we dealt with a real $2n$-dimensional vector space.

In QM the Hilbert space $\mathbb{H}$ is given as a vector space
over the field of complex numbers. Now we assume that two
Hermitian structures are given on it, that we will denote as
$(.,.)_{1}$and $(.,.)_{2}$ (both linear, for instance, in the
second factor). As in the real case, we look for the group that
leaves invariant both structure, that is the group of unitary
transformations w.r.t. both Hermitian structures. We call them
''bi-unitary''.

 In order to assure that $(.,.)_{1}$and \ $(.,.)_{2}$ do not define
different topologies on $\mathbb{H}$ it is necessary that there exists
$A,B\in\mathbb{R}$ , $0<A,B$ such that:%
\begin{equation}
A\left\|  x\right\|  _{2}\leq\left\|  x\right\|  _{1}\leq B\left\|  x\right\|
_{2}\ \ \ \ \ ,\forall x\in\mathbb{H}\
\end{equation}

The use of Riesz theorem on bounded linear functionals immediately implies
that there exists an operator $F$ defined implicitly by the equation:%
\begin{equation}
(x,y)_{2}=(Fx,y)_{1}\ \ ,\ \ \ \forall x,y\in\mathbb{H}\
\end{equation}

$F$ replaces the previous $G$ and $T$ tensors of the real vector space
situation, i.e. now it contains both the real and imaginary part of the
Hermitian structure and, in fact:%
\begin{equation}
F=(g_{1}+i\omega_{1})^{-1}\circ(g_{2}+i\omega_{2})
\end{equation}
It is trivial to show that $F$ is bounded, positive, and self-adjoint with
respect to both hermitian structures and that:%
\begin{equation}
\frac{1}{B^{2}}\leq\left\|  F\right\|  _{1}\leq\frac{1}{A^{2}}\, ;
\,\, \frac{1}{B^{2}}\leq\left\|  F\right\|
_{2}\leq\frac{1}{A^{2}}\, .
\end{equation}
If $\mathbb{H}$ is finite-dimensional, \ $F$ can be diagonalized, the two
hermitian structures decompose in each eigenspace of $F$, where they are
proportional and we get immediately that the group of \ bi-unitary
transformations is indeed:%
\begin{equation}
U(n_{1})\times U(n_{2})\times...\times U(n_{k})\ \ \ ,\ \ n_{1}+n_{2}%
+...n_{k}=n=\dim\mathbb{H}%
\end{equation}
where $n_{i}$ denotes the degeneracy of the $i-th$ eigenvalue of $F$.

In the infinite dimensional case $F$ may have a point part of the spectrum and
a continuum part. From the point part of the spectrum one gets $U(n_{1})\times
U(n_{2})\times...$ where now $n_{i}$ can be also $\infty$ . The continuum part
is more delicate to discuss. It \ will contain for sure the commutative group
$U_{F}$ of bi-unitary operators of the form $\left\{  e^{if(F)}\right\}  $
where $f$ is any real valued function (with very mild properties\cite{RS}).

The concept of genericity in the infinite dimensional case can not be given as
easily as in the finite dimensional case. One can say that the eigenvalues
should be non degenerate but what for the continuous spectrum? We give here an
alternative definition that works for the finite and infinite case as well.

Note first that any bi-unitary operator must commute with $F$.
Indeed:
$(x,U^{\dagger}FUy)_{2}=(Ux,FUy)_{2}=(FUx,Uy)_{2}=(Ux,Uy)_{1}=(x,y)_{1}%
=(Fx,y)_{2}=(x,Fy)_{2}$, from this: $U^{\dagger}FU=F$ \ ,
$[F,U]=0$ .

The group of \ bi-unitary operators therefore belongs to the commutant
$F^{\prime}$ of the operator $F.$ The genericity condition can be restated in
a purely algebraric form as follows:

\bigskip

\textbf{Definition: }Two hermitian forms are in a generic position iff
$F^{^{\prime\prime}}=F^{\prime},$ i.e. the bicommutant of $F$ coincides with
the commutant of $F.$

\bigskip

In other words this means that $F$ generates a complete set of observables.

This definition reduces, for the case of a pure point spectrum, to the
condition of nondegeneracy of the eigenvalues of $F$ and, in the real case, to
the minimum possible degeneracy of the eigenvalues of $T$ and $G$, that is two.

To grasp how the definition works, we will give some simple examples. Consider
: $(F\psi)(x)=x^{2}\psi(x)$ \ on the space $L_{2}([-b,-a]\cup\lbrack a,b])$
with $0<a<b$: then the operator $x$ , its powers $x^{n}$ and the parity
operator $P$ $\ $belong to$\ F^{\prime}$ while $F^{^{\prime\prime}}$ does not
contain $x$ (and any odd power of $x)$ because they do not commute with $P.$
So if $F=x^{2}$ the two hermitian structure are not in a generic position
because $F^{^{\prime\prime}}\subset$ $F^{\prime}$. On the contrary, on the
space $L_{2}([a,b]),$ $F^{^{\prime\prime}}=$ $F^{\prime}$ because a parity
operator $P$ does not exist in this case, so the two hermitian structure are
now in a generic position. In this case the group of bi-unitary operators is
$\left\{  e^{if(x^{2})t}\right\}  $ for the appropriate class of functions
$f$. In some sense, when a continuous part of the spectrum is considered,
there appears a continous family of $U(1)$'s as a counterpart of the discrete
family of $U(1)$'s corresponding to the discrete part of the spectrum.

\bigskip

\textbf{Remarks.}

i) \ Suppose that complex Hilbert spaces with two Hermitian structures have
been constructed from a given real vector space $V$ using two compatible and
admissible triples $(g_{1},\omega_{1},J_{1})$ and $(g_{2},\omega_{2},J_{2})$.
Then, by complexification, we get two different Hilbert space, each one with
its proper multiplication by complex numbers and with its proper Hermitian
structure. The previous case we have just studied is obtained if we assume
$J_{1}=J_{2}$. it is easy to show that this is a sufficient condition for
compatibility. This is the reason why in the quantum-mechanical case the group
of bi-unitary transformations is never empty, and the compatibility condition
is encoded already in the assumptions.

\bigskip

ii) \ If $J_{1}\neq J_{2}$ but the compatibility condition still holds, we
know that $V$ splits into $V_{+}\oplus V_{-}$, where $J_{1}=\pm J_{2}$ on
$V_{\pm}$ respectively. On $V_{+}$ we have the previous case, while on $V_{-}$
we get two Hermitian structures, one $\mathbb{C}$-linear and one
anti-$\mathbb{C}$-linear in the second factor (which one is which depending on
the complexification we have decided to use). From the point of view of the
group of unitary transformations, this circumstance is irrelevant, because the
set of unitary transformations does not change from being defined w.r.t. an
hermitian structure or to w.r.t. its complex conjugate. We conclude from this
that our analysis goes through in general, provided the compatibility
condition holds.

\bigskip

\section{\bigskip Conclusions.}

We will try now to summarize our main result, by restating it at
the same time in a more concise group-theoretical language. What
we have shown is, to begin with, that once two admissible triples:
$(g_{1},\omega_{1},J_{1})$ and \ $(g_{2},\omega_{2},J_{2})$ are
given on a real, even-dimensional vector space
$V\approx\mathbb{R}^{2n}$, they define two $2n$-dimensional real
representations, $U_{r}(2n;g_{1},\omega_{1})$ and \
$U_{r}(2n;g_{2},\omega _{2})$ of the unitary group $U(n)$,
$U_{r}(2n;g_{a},\omega_{a})$ $(a=1,2)$ being the group of
transformations that leave simultaneously $g_{a}$ and
$\omega_{a}$ (and hence $J_{a}$) invariant. Their intersection :%
\begin{equation}
W_{r}=:\{U_{r}(2n;g_{1},\omega_{1})\cap U_{r}(2n;g_{2},\omega_{2})\}
\end{equation}
will be their common subgroup that is an invariance group for both triples.
The assumption of compatibility\footnote{As the previous two-dimensional
example shows explicitly, but it should be clear by now also in general.}
implies that $W_{r}$ should not reduce to the identity alone.

If the two triples are in a generic position, then:%
\begin{equation}
\underset{n\text{ \ \ }factors}{W_{r}=\underbrace{SO(2)\times SO(2)\times
...\times SO(2)}}%
\end{equation}
where, here: $SO(2)\approx U(1)$ or, more generally if the genericity
assumption is dropped:%
\begin{equation}
W_{r}=U_{r}(2r_{1};g,\omega)\times U_{r}(2r_{2};g,\omega)\times...\times
U_{r}(2r_{k};g,\omega)
\end{equation}
where: $r_{1}+r_{2}+...+r_{k}=n$ and $(g,\omega)$ is any one of the two pairs
$(g_{1},\omega_{1})$ or $(g_{2},\omega_{2})$.

The real vector space $\ V\approx\mathbb{R}^{2n}$ will decompose then into a
direct sum of even-dimensional subspaces that are mutually orthogonal w.r.t.
both metrics, and on each subspace the corresponding (realization of the)
special orthogonal group will act irreducibly.

Alternatively, we can complexify $V\approx\mathbb{R}^{2n}$, and that in two
different ways, using the two complex structures that are at our disposal. The
equivalent statement in the complex framework will be then:

Given two hermitian structures $h_{a},a=1,2$ on a complex $n$-dimensional
vector space $\mathbb{C}^{n}$, they define two representations $U(n;h_{a}%
),a=1,2$ of the group $U(n)$ on the same $\mathbb{C}^{n}$. $U(h_{1},n)$ (resp.
$U(h_{2},n)$) will be the group of transformations that are unitary with
respect to $h_{1}$ (resp. $h_{2}$). The group $W$ of simultaneous invariance
for both hermitian structures :
\begin{equation}
W\equiv\left\{  U(h_{1},n)\cap U(h_{2},n)\right\}  \ \ \ \ \ .
\end{equation}
will be a subgroup of both $U(h_{1},n)$ and $U(h_{2},n)$, and our assumption
of compatibility of $\ $the $\ h_{a}$ 's \ implies that the component of $W$
connected to the identity should not reduce to the identity alone.

The assumption of genericity implies that:%
\begin{equation}
\underset{n\text{ \ \ }factors}{W=\underbrace{U(1)\times U(1)\times...\times
U(1)}} \label{tes}%
\end{equation}
\newline If the assumption of genericity is dropped, one can easily show,
along the same lines as in the generic case, that $W$ will be of the form:
\begin{equation}
W=U(r_{1})\times U(r_{2})...\times U(r_{k})\ \ \ , \label{des}%
\end{equation}
with $r_{1}+r_{2}+...+r_{k}=n$ . $\mathbb{C}^{n}$ will decompose accordingly
\ into a direct sum of subspaces that will be mutually orthogonal \ with
respect to both $h_{a}$ 's, and on each subspace the appropriate $U(r)$ will
act irreducibly.

\bigskip We have also shown that these results generalize to the
infinite-dimensional case as well. Some extra assumptions must be added on the
Hermitian structures in order that they define the same topology in
$\mathbb{H}$ and an appropriate definition of genericity must also be given.
Then, a decomposition like in Eqns.(\ref{tes}) and (\ref{des}) is obtained,
possibly with denumberable discrete terms and a continuum part as well. We
note that, in the spirit of this work where two Hermitian structures are given
from the very beginning, it is natural to supplement the compatibility
condition, in the infinite-dimensional case, with a topological equivalence
condition. However from the point of view of the study of bi-hamiltonian
systems, where a fixed dynamics is given, it would be more natural to assume
some weaker regularity condition, for instance that the given dynamics should
be continuous with respect to both structures.

Recently, bi-Hamiltonian systems ''generated'' out of a pencil of compatible
Poisson structures have been considered \cite{Gelf}, also in connection with
the separability problem \cite{ibort}. It should be noticed that our
compatible structures would give rise to a pencil of compatible triples
defined by:
\begin{equation}
g_{\gamma}=g_{1}+\gamma g_{2}\ \ ,\ \ \omega_{\gamma}=\omega_{1}+\gamma
\omega_{2}\ ,\ J_{\gamma}\
\end{equation}

A systematic comparison with this approach is presently under consideration.

\end{document}